\documentclass[12pt]{article}
\usepackage{graphics}
\usepackage{color}
\providecommand{\SU}[1]{}
\renewcommand{\SU}[1]{\ensuremath{\mathrm{SU}(#1)}}

\providecommand{\Dbrane}[1]{}
\renewcommand{\Dbrane}[1]{\ensuremath{D#1\mathrm{-brane}}}
\providecommand{\Dbranes}[1]{}
\renewcommand{\Dbranes}[1]{\ensuremath{D#1\mathrm{-branes}}}
\providecommand{\vev}{}
\renewcommand{\vev}{vev}
\providecommand{\vevs}{}
\renewcommand{\vevs}{vevs}

\providecommand{\Kahler}{}
\renewcommand{\Kahler}{K\"ahler}
\providecommand{\Fbar}{}
\renewcommand{\Fbar}{\overline{F}}

\begin{document}
\baselineskip 24pt

\newcommand{\sheptitle}
{Canonical normalisation and Yukawa matrices}
\newcommand{\shepauthor}
{S. F. King and I. N. R. Peddie}
\newcommand{\shepaddress}
{School of Physics and Astronomy, University of Southampton \\
  Southampton, SO17 1BJ, U.K.}
\newcommand{\shepabstract}
{
We highlight the important r\^{o}le that 
canonical normalisation of kinetic terms in flavour models 
based on family symmetries can play
in determining the Yukawa matrices. Even though the
kinetic terms may be correctly canonically normalised
to begin with, they will inevitably be driven into 
a non-canonical form by a similar operator expansion
to that which determines the Yukawa operators.
Therefore in models based on family symmetry
canonical re-normalisation is mandatory before the
physical Yukawa matrices can be extracted.
{\em In nearly all examples in the literature this is not done.}
As an example
we perform an explicit calculation of such mixing associated
with canonical normalisation of the \Kahler{} metric in a
supersymmetric model based on $SU(3)$ family symmetry,
where we show that such effects can significantly change the form of the
Yukawa matrix. In principle quark mixing 
could originate entirely from canonical normalisation,
with only diagonal Yukawa couplings before canonical normalisation.
}

\begin{titlepage}
  \begin{flushright}
    hep-ph/0312237 \\
    SHEP/0334
  \end{flushright}
  \begin{center}
    {\large{\bf \sheptitle}} \\
    \shepauthor \\
    \mbox{} \\
    {\it \shepaddress } \\
    {\bf Abstract } \\
     \bigskip    
  \end{center}
\setcounter{page}{0}
\shepabstract
\begin{flushleft}
  \today
\end{flushleft}
\end{titlepage}

\newpage

\section{Introduction}
\label{sec:introduction}

There is great interest in the literature in trying to understand the
hierarchical pattern of Standard Model fermion masses, the smallness
of the quark mixing angles and the two large and one small neutrino
mixing angles.
One popular way of doing this is to extend either the
Standard Model, or one of its more common supersymmetric extensions, by
adding a gauge or global family symmetry, $G_F$ which is subsequently
broken \cite{Froggatt:1978nt}.

In such models based on family symmetry $G_F$, Yukawa
couplings arise from Yukawa operators which are typically
non-renormalisable and involve extra heavy
scalar fields, $\phi$, coupling to the usual three fields, for example:
\begin{equation}
  \label{eq:34}
  \mathcal{O}_Y = F \overline{F} H \left(\frac{\phi}{M}\right)^n
\end{equation}
where $F$ represents left-handed fermion fields, $\overline{F}$
represents the $CP$-conjugate of right-handed fermion fields, $H$
represents the Higgs field, and M is a heavy mass scale which acts as
an ultraviolet (UV) cutoff.  In the context of supersymmetric (SUSY)
field theories, all the fields become superfields.  The operators
in Eq.\ref{eq:34} are invariant under $G_F$, but when the scalar
fields $\phi$ develop vacuum expectation values (vevs) the family
symmetry is thereby broken and the Yukawa couplings are generated. The
resulting Yukawa couplings are therefore effective couplings expressed
in terms of an expansion parameter, $\epsilon$, which is the ratio of
the \vev{} of the heavy scalar field to the UV cut-off, $\epsilon =
\frac{\left<\phi\right>}{M}$.  Explaining the hierarchical form of the
Yukawa matrices then reduces to finding an appropriate symmetry 
$G_F$ and field content which leads to acceptable forms of
Yukawa matrices, and hence fermion masses and mixing angles,
at the high energy scale. 

Over recent years there has been a huge activity in this 
family symmetry and operator approach to 
understanding the fermion masses and mixing angles \cite{Ross:2000fn},
including neutrino masses and mixing angles
\cite{King:2003jb}.
However, as we shall show in this paper, in analysing such
models it is important to also consider the corresponding
operator expansion of the kinetic terms. The point is that,
even though the kinetic terms may be correctly canonically
normalised to begin with, they will inevitably be driven to 
a non-canonical form by a similar operator expansion to that
which determines the Yukawa operator. In order to extract
reliable predictions of Yukawa matrices, it is mandatory to 
canonically re-normalise the kinetic terms once again before
proceeding. {\em In nearly all examples in the literature this
is not done.} The main point of our paper is thus to highlight
this effect and to argue that it is sufficiently important that 
it must be taken into account before
reliable predictions can be obtained.

Many approaches combine the family symmetry and operator approach
with supersymmetric grand unified theories (SUSY GUTs)
\cite{Ross:2000fn,King:2003jb}. Such models tend
to be more constraining, because the Yukawa matrices at the high scale
should have the same form, up to small corrections from the breaking
of the unified symmetry. The same comments we made above 
also apply in the framework of SUSY GUTs. In the SUSY case the 
Yukawa operators arise from the
superpotential $W$, and 
the kinetic terms and scalar masses, as well as gauge interaction
terms come from the \Kahler{} potential, $K$. 
In nearly all examples in the literature
the superpotential $W$ has been analysed independently of the
\Kahler{} potential, $K$, leading to most of the published
results being inconsistent. The correct procedure which should
be followed is as follows.

To be consistent, the \Kahler{} potential, $K$, 
should also be written down to the same
order $M^{-n}$ as the superpotential $W$. Having done this, one
should proceed to calculate the elements of the \Kahler{} metric,
$\tilde{K}_{ij}$, which are second derivatives with respect to fields
of the \Kahler{} potential $\tilde{K}_{i\overline{j}} =
\frac{\partial^2 K}{\partial \phi_i \partial \phi^\dag_j}$. However,
in order to have canonically normalised kinetic terms, the \Kahler{}
metric has to itself be canonically normalised
$\tilde{K}_{i\overline{j}} = \delta_{i\overline{j}}$. In making this
transformation, the superfields in the \Kahler{} potential are first
being mixed and then rescaled.  Once this has been done, the
superfields in the superpotential must be replaced by the canonically
normalised fields.

Canonical normalisation is not of course a new invention,
it has been known since the early days of supergravity 
\cite{Soni:1983rm}. However, as we have mentioned, for some
reason this effect has been largely ignored in the model building 
community. A notable exception is the observation some 
time ago by Dudas, Pokorski and Savoy \cite{Dudas:1995eq},
that the act of canonical normalisation will change the Yukawa
couplings, and could serve to cover up ``texture zeros'', 
which are due to an Abelian family symmetry which does not allow
a specific entry in the Yukawa matrix and is therefore
manifested as a zero at high energies. This issue has been resurrected
for abelian family models recently \cite{Jack:2003pb}.
However, as we have already
noted, this observation has not been pursued or developed
in the literature, but instead has been largely ignored.

In this paper we consider the issue of canonical
normalisation in the framework of non-Abelian symmetries, in which the
Yukawa matrices are approximately symmetric. 
In such a framework we show that the effects of canonical
normalisation extend beyond the filling in of ``texture zeros'',
and can also change the expansion order of the leading 
non-zero entries in the Yukawa matrix. As an example
we perform an explicit calculation of such mixing associated
with canonical normalisation of the \Kahler{} metric in a recent
supersymmetric model based on $SU(3)$ family symmetry
where we show that such effects can significantly change the form of the
Yukawa matrix. The $SU(3)$ model we consider is
a grossly simplified version of the realistic model in
\cite{King:2003rf}, where we only consider the case of a single
expansion parameter and perform 
our calculations in the 23 sector of the theory
for simplicity, although we indicate how the results
can straightforwardly be extended to the entire theory.
An alternative scenario in which in principle quark mixing could originate 
entirely from canonical normalisation,
with only diagonal Yukawa couplings before canonical normalisation,
is also discussed.

The outline of the rest of the paper is as follows. In
section~\ref{sec:general-overview} we discuss the issues
surrounding canonical normalisation in the Standard Model
supplemented by a family symmetry, first without then with SUSY.
In the SUSY case we discuss the scalar mass squared and Yukawa
matrices for two types of \Kahler{} potentia where only one superfield
contributes to supersymmetry breaking. 
In section~\ref{sec:toy-model} we discuss a particular 
model in some detail as a concrete example,
namely the simplified $SU(3)$ family symmetry model,
focusing on the second and third generations of matter, 
later indicating how the results can be extended to 
all three families. We conclude in section~\ref{sec:conclusions}.

\section{Canonical normalisation}
\label{sec:general-overview}

\subsection{Standard Model with a Family Symmetry}
\label{sec:standard-model-with-family-sym}

In this section we first consider
extending the Standard Model gauge group
with a family symmetry, under which each generation has a different
charge ( for abelian family symmetries ) or representation ( for
non-abelian family symmetries ). The family symmetry typically
prohibits renormalisable Yukawa couplings (except
possibly for the third family) but allows non-renormalisable
operators, for example:
\begin{equation}
  \label{eq:24}
  \mathcal{O}_Y = F^i H \overline{F}^j \frac{\phi_i \phi_j}{M^2}
\end{equation}
where $i,j$ are generation indices,
$M$ is some appropriate UV cutoff, $F$ represents left-handed fermion fields,
and $\overline{F}$ represents $CP$-conjugates of right-handed
fermion fields, and $H$ is a Higgs field. When 
the flavon scalar field $\phi$ gets a \vev{},
which breaks the family symmetry, effective Yukawa couplings
are generated:
\begin{equation}
  \label{eq:25}
  Y_{ij} = \frac{\left<\phi_i\right>\left<\phi_j\right>}{M^2}
\end{equation}
The effective Yukawa matrices are determined by the operators allowed by
the symmetries of the model,
$G_F\otimes SU(3)_c \otimes SU(2)_L \otimes U(1)_Y$, 
as well as the form that the vev of $\phi$ takes.

Even though the kinetic terms are correctly canonically 
normalised to begin with, they will receive non-renormalisable
corrections arising from operators allowed by the family symmetry,
which will cast them into non-canonical form. For example,
\begin{equation}
  \label{eq:35}
  F^\dag_i \,/\!\!\!\partial\,  F^j \left( \delta^i_j  + \frac{\phi^i \phi^\dag_j }{M^2} \right)
\end{equation}
This leads to a non-canonical kinetic term when $\phi$ is replaced by its \vev.
It is therefore mandatory to perform a further canonical
re-normalisation of the kinetic terms, before analysing the physical
Yukawa couplings. The canonical normalisation amounts to 
a transformation which is not unitary but which gives all the fields
canonical kinetic terms. 
The kinetic part of a theory with a Higgs scalar field $H$, 
a fermionic field $F^i$
and the field strength tensor $F^{\mu\nu}$ corresponding to a gauge field $A^\mu$ when canonical will
look like:
\begin{equation}
  \label{eq:29}
  \mathcal{L}_{\mathrm{canonical}} = \partial_\mu H \partial^\mu H^* 
  + F^\dag_i \,/\!\!\!\partial\, F^i - \frac{1}{4} F^{\mu\nu}F_{\mu\nu}
\end{equation}
Once we have done this normalisation, we have to rewrite all of our
interactions in terms of the canonical fields with the shifted fields.

The important point we wish to emphasise is that all the interaction terms
should be expressed in terms of canonical fields, before making
any physical interpretation. If this is not done, as is often the
case in the literature, then the results will not be reliable.

\subsection{SUSY Standard Model with Family Symmetry}
\label{sec:susy-standard-model-with-family-sym}

In the context of supersymmetric theories,
it turns out to be possible
to automatically canonically normalise all the fields in the theory at
once. However these transformations are not always
simple, and in practice calculating the relevant transformations may well turn
out to be intractable for any given model.

The aim of SUSY model builders with respect to flavour is
two-fold. The primary wish is to generate a set of effective Yukawa
matrices which successfully predict the quark and lepton masses and
mixing angles as measured by experiment. However, because of the
parameters associated with softly broken SUSY models, there exist
dangerous one-loop diagrams which lead to processes such as
$b\rightarrow s\gamma$ and $\mu\rightarrow e\gamma$ at rates much
greater than predicted by the Standard Model and also much greater
than measured by experiment. A successful SUSY theory of flavour
will therefore successfully describe fermion masses and 
mixing angles, while simultaneously controlling such flavour changing
processes induced by loop diagrams involving
sfermion masses which are off-diagonal in the basis
where the quarks and leptons are diagonal.

In a SUSY or supergravity (SUGRA) model, very often the starting
point in addressing the flavour problem is to propose a 
set of symmetries that will give rise to non-renormalisable
superpotential operators which will lead to a hierarchical form for
our Yukawa matrices, arising from some effective Yukawa operators 
as discussed previously.
Extra fields, $\phi$ are introduced that
spontaneously break the extra family symmetries.  
The general form of the superpotential is :
\begin{equation}
  \label{eq:19}
  W = F^i \overline{F}^j H w_{ij}\left({\phi}/{M} \right)
\end{equation}
Here $w_{ij}(\phi /M)$ is a general function of the extra fields, $\phi$, which has mass
dimension zero and contracts with $F^i \overline{F}^j$ to make $W$ a singlet
of the extended symmetry group. 

In models of this type, the amount of flavour violation is
proportional to the size of the off-diagonal elements in the scalar
mass matrices at the electroweak (EW) scale when the scalar mass
matrices have been rotated to the basis where the Yukawas are diagonal
(the super-CKM basis). Since the quark mixing angles are small, this
suggests that any large scalar mixings at the electroweak scale would
remain large when in the super-CKM basis. Since we would generally not
expect the RG running of the scalar mass matrices to tune large
off-diagonal values to zero, we would expect to be in trouble
if there are large off diagonal scalar mass mixings predicted at the
high energy scale. This scale might be, for example, the unification
scale in a SUSY GUT.

We now proceed to outline the argument that this will not be a problem
in general terms in the simplest case, gravity-mediated supersymmetry
breaking, where the breaking is due to a single hidden sector
superfield, S. As examples, 
we shall consider the \Kahler{} potential in two forms.
The first form is:
\begin{equation}
  \label{eq:2}
  K_1 = \ln(S + \overline{S}) + F^i F^\dag_j k^j_i\left({\phi}/{M}\right) 
  + \overline{F}^i \overline{F}^\dag_j  \overline{k}^j_i 
\left({\phi}/{M}\right)
\end{equation}
The second form we consider is:
\begin{equation}
  \label{eq:3}
  K_2 = \frac{S\overline{S}}{M^2}
\left( F^i F^\dag_j k^j_i\left({\phi}/{M}\right) 
+ \overline{F}^i \overline{F}^\dag_j 
\overline{k}^j_i \left({\phi}/{M}\right) \right)
\end{equation}
Here $k(\phi)$ and $\overline{k}(\phi)$ represent functions of the
various $\phi$ fields that can be contracted with the matter fields to
make the \Kahler{} potential a singlet and of the correct mass
dimension.

Since we are looking at gravity-mediated SUSY breaking, we may use the
SUGRA equations which relate the {\em non-canonically normalised}
soft scalar mass squared matrices 
$m^2_{\overline{a}b}$ in the soft SUSY breaking
Lagrangian to the \Kahler{} metric $\tilde{K}_{\overline{a}b} =
\frac{\partial^2 K}{\partial \phi^\dag_a \partial \phi^b}$, and the
vevs of the auxiliary fields which are associated with the
supersymmetry breaking, $F_m$ \cite{Soni:1983rm}:
\begin{equation}
  \label{eq:4}
  m^2_{\overline{a}b} = m_{3/2}^2  \tilde{K}_{\overline{a}b} 
  - F_{\overline{m}} 
  \left(
    \partial_{\overline{m}}\partial_n \tilde{K}_{\overline{a}b}
    - \partial_{\overline{m}} \tilde{K}_{\overline{a}c} 
    (\tilde{K}^{-1})_{c\overline{d}} \partial_n \tilde{K}_{\overline{d}b} 
  \right) F_n
\end{equation}
where we have assumed a negligibly small cosmological constant.
Roman indices from the middle of the alphabet are taken to be over the
hidden sector fields, which in our case can only be the singlet field
S associated with SUSY breakdown. As it happens, for both $K_1$ and
$K_2$, the {\em non-canonically normalised} mass matrix reduces to:
\begin{equation}
  \label{eq:5}
  m^2_{\overline{a}b} = m_{3/2}^2 \tilde{K}_{\overline{a}b}
\end{equation}
This is obvious for $K_1$, since the \Kahler{} metric doesn't involve
S, so partial derivatives with respect to S will give zero. To see
that eq.~(\ref{eq:4}) reduces to eq.~(\ref{eq:5}) for $K_2$, is less
obvious. We first write:
\begin{equation}
  \label{eq:22}
  \tilde{K}_{\overline{a}b} = \frac{S\overline{S}}{M^2} \mathcal{M}_{\overline{a}b}
\end{equation}
Substituting this into eq.~(\ref{eq:4}) gives a non-canonically
normalised scalar mass squared matrix:
\begin{equation}
  \label{eq:6}
  m^2_{\overline{a}b} = m_{3/2}^2 \tilde{K}_{\overline{a}b}
  - F_{\overline{S}}\left(
    \frac{1}{M^2} \mathcal{M} - \frac{S}{M^2}\mathcal{M} \frac{M^2}{S\overline{S}} \mathcal{M}^{-1}
    \mathcal{M}\frac{\overline{S}}{M^2} \right) F_S
\end{equation}
It is clear that eq.~(\ref{eq:6}) reduces to eq.~(\ref{eq:5}).
However, the physical states are those for which the \Kahler{} metric
is canonically normalised, $\tilde{K} = 1$.  This is attained by
$\tilde{P}^\dag \tilde{K} \tilde{P} = 1$. In order to canonically
normalise the mass matrix, we apply the same transformation, and
find that the {\em canonically normalised} squark mass squared matrix
then takes the universal form:
\begin{equation}
  \label{eq:7}
  m^2_{\mathrm{c.n.}} = m_{3/2}^2 \mathbf{1}
\end{equation}

We conclude that models with \Kahler{} potentials like $K_1$ or $K_2$
will result in universal sfermion masses 
at the high-energy scale. Of course all this is well known,
and it has long been appreciated that this would tame
the second part of the flavour problem, flavour violating
decays.  However, what is less well appreciated at least amongst the 
model building community, is that 
canonical normalisation corresponds to redefining
the fields in the \Kahler{} potential, and one must therefore also
redefine these fields in the same way in the superpotential. 
Unless this is done consistently it could lead to a
problem with the first part of the flavour problem, because the
shifted fields may well no longer lead to a phenomenologically
successful prediction of the masses and mixing angles for the quarks
and leptons.

The procedure is easily formalised in SUSY theories by 
writing all the fields as $\psi_i$, $i = 1 \cdots n$.
Then, at least in principle, if not in practice, a matrix, $P$ 
can be found which
transforms the vector $\mathbf{\psi}$ to the vector
$\mathbf{\psi}_c$ for which the \Kahler{} metric is canonically
normalised:
\begin{equation}
  \label{eq:20}
  \mathbf{\psi} \rightarrow \mathbf{\psi}_c = P \cdot \mathbf{\psi}
\end{equation}
This procedure can be followed providing 
$\mathbf{\psi}$ if $P$ is not a singular
matrix. Having found $P$, one we can write $\mathbf{\psi} = P^{-1} \mathbf{\psi}_c$.
We can then substitute these into the superpotential
now expressed in terms of fields which correspond to 
a correctly canonically normalised \Kahler{} metric:
\begin{equation}
  \label{eq:21}
  W^\prime = F^{\prime\,i} \overline{F}^{\prime\,j} H^\prime w^\prime_{ij} 
  \left( {\phi^\prime}/{M} \right)
\end{equation}

It should be noted that despite the shifts originating from the \Kahler{} potential,
which is a non-holomorphic function of the fields, the shifted fields only
depend on the \vevs{} of the flavon and its hermitian conjugate. Therefore,
the shifted superpotential will remain holomorphic in terms of fields. That
is to say, the shifted fields will be a function of the corresponding unshifted field,
and the \vevs{} which break the family symmetry. Here $\theta$ represents any field,
$\left<\phi\right>$ the \vev{} of a flavon field, and $\left<\phi^\dag\right>$ the \vev{}
of the hermitian conjugate of a flavon field:
\begin{equation}
  \label{eq:26}
  \theta \rightarrow \theta^\prime\left(\theta, \left<\phi\right>, \left<\phi^\dag\right>\right)
\end{equation}

At this point, if we had a specific model, we would then need to check 
that the Yukawas are viable. This is then the correct procedure which 
must be followed in analysing a general model. We now turn to 
a particular example which illustrates the effects described above,
in the framework on a non-Abelian family symmetry.

\section{A SUSY Model based on $SU(3)$ family symmetry}
\label{sec:toy-model}

\subsection{The quark sector}
\label{sec:quark-sector-ncn}

As an example of the general considerations above, and in order to
determine the quantitative effects of canonical normalisation, 
we now turn to a particular example based on
$\SU{3}_F$ family symmetry.
As mentioned the model we consider is a simplified version of the
realistic model by King and Ross \cite{King:2003rf}
in which we assume only a single expansion parameter. 
For simplicity, we shall also ignore the presence of the first
generation, although later we shall indicate how the results
may be extended to the three family case. This model
is based on a SUSY Pati-Salam model with a gauged \SU{3}
family symmetry extended by a $Z_2 \otimes U(1)$ global symmetry. 
As shown in Table 1, the left-handed matter is contained in
$F^i$, the right-handed
matter is contained in a left-handed field $\overline{F}^i$ . 
The MSSM Higgs doublets are contained in H; $\Sigma$ is a field which
has broken $SO(10)$ to $SU(4)_{PS} \otimes SU(2)_L \otimes SU(2)_R$.
There are two $SU(3)_F$-breaking fields, $\phi_3$ and $\phi_{23}$.

\begin{table}[htbp]
  \centering
  \begin{tabular}{|c|c|c|c|c|c|c|}
    \hline
    Field & $SU(3)_F$ & $SU(4)_{PS}$ & $SU(2)_L$ & $SU(2)_R$ & $Z_2$ & $U(1)$ \\
    \hline
    $F$ & $3$ & $4$ & $2$ & $1$ & + & 0 \\
    $\overline{F}$ & $3$ & $\overline{4}$ & $1$ & $2$ & + & 0 \\
    $H$ & $1$ & $1$ & $2$ & $2$ & + & 8 \\
    $\Sigma$ & $1$ & $15$ & $1$ & $1$ & + & 2 \\
    $\phi_3$ & $\overline{3}$ & $1$ & $1$ & $1$ &-& $-4$ \\
    $\phi_{23}$ & $\overline{3}$ & $1$ & $1$ & $1$ &+& $-5$\\
    \hline
  \end{tabular}
  \caption{The field content of the toy model}
  \label{tab:field_content}
\end{table}

The superpotential has to be a singlet under the combined gauge group 
$SU(4)_{PS} \otimes SU(2)_L \otimes SU(2)_R \otimes SU(3)_F$ and also
neutral under $Z_2 \otimes U(1)$. Because of this,
the standard Yukawa superpotential:
\begin{equation}
  \label{eq:standard_yukawa_superpotential}
  W = F^i \overline{F}^j H Y_{ij}
\end{equation}
is not allowed because of the $Z_2
\otimes U(1)$. As such, we have to move to a superpotential containing
non-renormalisable terms.  We view this as being the superpotential
corresponding to a supersymmetric effective field theory, where some
heavy messenger states and their superpartners have been integrated
out. Then, assuming that the messenger states have the same
approximate mass scale, we write:
\begin{equation}
  \label{eq:toy_model_superpotential}
  W = \frac{1}{M^2} a_1 F^i \overline{F}^j H \phi_{3,i}\phi_{3,j}
  + \frac{1}{M^3} a_2 F^i \overline{F}^j H \Sigma \phi_{23,i} \phi_{23,j}
\end{equation}
The $a_i$ are parameters that are expected to be of the order of
unity, $M$ is the appropriate UV cutoff of the effective field
theory. This will clearly lead to a set of effective Yukawa terms when
the fields $\phi_3$ and $\phi_{23}$ gain \vevs{} which break the
family symmetry.

We choose the vacuum structure after King and Ross \cite{King:2003rf}:
\begin{equation}
  \label{eq:vevs}
  \left< \phi_3 \right> = \left( \begin{array}{c} 0 \\ 1 \end{array} \right ) a
\; ; \;
\left< \phi_{23} \right> = \left( \begin{array}{c} 1 \\ 1 \end{array} \right) b \; ;
\;
\left<\Sigma \right> = \sigma
\end{equation}
And we then trade these for a single expansion parameter, $\epsilon
\approx \frac{1}{10}$.
\begin{equation}
  \label{eq:vevs_as_epsilons}
  \frac{a}{M} = \sqrt{\epsilon} \; ; \; \frac{b}{M} = \epsilon \; ; \; \frac{\sigma}{M} = \epsilon
\end{equation}
Substituting eqs.~(\ref{eq:vevs},~\ref{eq:vevs_as_epsilons}) into
eq.~(\ref{eq:toy_model_superpotential}), we can write down our
high-energy Yukawa matrix:
\begin{equation}
  \label{eq:1}
  Y_{\mathrm{n.c.}} = 
  \left(
    \begin{array}{cc}
      a_2 \epsilon^3 & a_2 \epsilon^3 \\
      a_2 \epsilon^3 & a_1 \epsilon + a_2 \epsilon^3 
    \end{array}
  \right)
\end{equation}
We write it as $Y_{\mathrm{n.c.}}$ to represent the fact that it is
the Yukawa matrix corresponding to the non-canonical \Kahler{} metric.

\subsection{The squark sector}
\label{sec:squark-sector-ncn}

In order to write down the squark mass matrices, the first step is to
write down our \Kahler{} potential. This should be the most general
\Kahler{} potential consistent with the symmetries of our model up to
the same order in inverse powers of the UV cutoff as the
superpotential is taken to. In our case, this is $M^{-3}$. However,
from the general arguments of section \ref{sec:general-overview}, we know that
if we pick our \Kahler{} potential, $K$ to be of the form as $K_1$
(eq.~(\ref{eq:2})) or $K_2$ (eq.~(\ref{eq:3})) then we will have
universal scalars.

The non-canonical form of the scalar mass-squared matrix is:
\begin{equation}
  \label{eq:30}
  m^2_{\mathrm{n.c.}} \sim 
  \left(
    \begin{array}{cc}
      1 + \epsilon & \epsilon^2 \\
      \epsilon^2 & 1 + \epsilon
    \end{array}
  \right)
\end{equation}

However, we already know exactly what the canonical form of this
matrix will look like:
\begin{equation}
  \label{eq:11}
  m^2 = m^2_{3/2} 
  \left(
    \begin{array}{cc}
      1 & 0 \\
      0 & 1
    \end{array}
  \right)
\end{equation}
This universal form is a direct result of the simple supersymmetry
breaking mechanism that we have and canonical normalisation,
and is independent of other details about the model.

\subsection{\Kahler{} potential for the model}
\label{sec:our-kahler-potential}

We saw in the previous subsection that we will not end up with
dangerous off-diagonal elements in the scalar mass matrices for
general \Kahler{} potentials of the type we are going to look at.
We must now write down our \Kahler{}
potential.  We choose this to be of the same form $K_1$. There will be
no $M^{-3}$ terms, so it will suffice to write this down up to
$\mathcal{O}(M^{-2})$.

The matrices which diagonalise the \Kahler{} metric will in general be
large and intractable.  In order to proceed, we will have to make
some simplifying assumptions. We first
assume that the \Kahler{} metric $\tilde{K}_{\overline{a}b} =
\frac{\partial^2 K}{\partial \phi^\dag_a \partial \phi^b}$
is block diagonal, of the form:
\begin{equation}
  \label{eq:8}
  \tilde{K}_{\overline{a}b} = 
  \left(
    \begin{array}{ccccc}
      \tilde{K}_{LH} \\
      & \tilde{K}_{RH} \\
      && \tilde{K}_{\phi} \\
      &&& \tilde{K}_\Sigma \\
      &&&& \tilde{K}_H 
  \end{array}
    \right)_{\overline{a}b}
\end{equation}
In this, $\tilde{K}_{LH}$ represents the block for chiral
superfields, $F$, containing left-handed matter; $\tilde{K}_{RH}$ represents chiral
superfields, $\overline{F}$, containing right-handed matter; $\tilde{K}_\phi$ represents the
$SU(3)_F$ breaking Higgs fields, $\phi_{23}$ and $\phi_3$;
$\tilde{K}_\Sigma$ represents the block for the Higgs field that break
the GUT symmetry down to the MSSM gauge group, $\Sigma$; finally, the
block $\tilde{K}_H$ represents the block corresponding to the MSSM
Higgs fields, $H$. The block diagonal assumption is equivalent to 
switching off some terms in the \Kahler{} potential. 
The remaining terms in the \Kahler{} potential
are listed below:
\begin{eqnarray}
  \nonumber
  K &&= \ln(S+\overline{S}) + b_0 F^i F^\dag_i + \frac{1}{M^2} F^i F^\dag_j  \left\{ 
    \phi_3^k \phi_{3,l}^\dag ( b_1 \delta^l_i \delta^j_k + b_2 \delta^j_i \delta^l_k ) \right.\\
  \nonumber
  && + \phi_{23}^k \phi_{23,l}^\dag ( b_3 \delta^l_i \delta^j_k + b_4 \delta_i^j \delta^l_k ) 
   \left. + b_5 HH^\dag \delta^j_i + b_6 \Sigma \Sigma^\dag \delta^j_i \right \} \\
  \nonumber
  && + c_0 \Fbar^i \Fbar^\dag_i + \frac{1}{M^2} \Fbar^i \Fbar^\dag_j 
  \left\{ \phi_3^k \phi^\dag_{3,l}(c_1 \delta^l_i \delta^j_k + c_2 \delta^j_i \delta^l_k )\right.
  \\
  \nonumber
  && \left. + \phi_{23}^k \phi_{23,l}^\dag ( c_3 \delta^l_k \delta^j_k + c_4 \delta^j_i \delta^l_k )
  + c_5 HH^\dag \delta^j_i + c_6 \Sigma \Sigma^\dag \delta^j_i \right\}
  \\
  \nonumber
  && + d_1 \phi_3^i \phi_{3,i}^\dag + d_2 \phi_{23}^i \phi_{23,i}^\dag + d_3 HH^\dag + d_4 \Sigma\Sigma^\dag \\
  \nonumber
  && + \frac{1}{M^2}\left\{
    \phi_3^i \phi_{3,j}^\dag \phi_3^k \phi_{3,l}^\dag d_5 \delta^j_i \delta^l_k 
    + \phi_3^i \phi_{3,j}^\dag \phi_{23}^k \phi_{23,l}^\dag ( d_6 \delta^j_i \delta^l_k
    + d_7 \delta^j_k \delta^l_i ) \right. \\
  && \left. + \phi_{23}^i \phi_{23,j}^\dag \phi_{23}^k \phi_{23,l}^\dag d_8 \delta^j_i \delta^l_k
    + d_9 H H^\dag HH^\dag + d_{10} \Sigma\Sigma^\dag\Sigma\Sigma^\dag \right\}
  \label{eq:9}
\end{eqnarray}

Having done this, we now need to calculate the \Kahler{} metric
$\tilde{K}$. But since we have set $K$ up specifically such that it is
block diagonal, we can instead work out the non-zero blocks,
$\tilde{K}_{LH}$, $\tilde{K}_{RH}$, $\tilde{K}_\phi$,
$\tilde{K}_\Sigma$ and $\tilde{K}_H$. Once we have done so, we need to
canonically normalise them. This is done in two stages. The first is a
unitary transformation to diagonalise each block $\tilde{K}_i$:
\begin{equation}
  \label{eq:10}
  \mathcal{L} \supset F^\dag\tilde{K}  F\rightarrow (F^\dag U) ( U^\dag \tilde{K}  U)(  U^\dag F )
  = F^\prime \tilde{K}^\prime F^{\dag\,\prime}
\end{equation}

The mixed \Kahler{} metric, $\tilde{K}^\prime$, is now diagonal. Then
we rescale the fields by a diagonal matrix R such that $R_i =
(\tilde{K}_i^\prime)^{-1/2}$. These new superfields are then
canonically normalised.

Then:
\begin{equation}
  \label{eq:28}
  \mathcal{L} \supset (F^\dag U R^{-1}) \underbrace{(R U^\dag \tilde{K} U R)}_{\mathbf{1}} (R^{-1} U^\dag F)
\end{equation}
If we call $P$ the matrix which converts $F$ to the canonical field $F_c$, then we can note two things. Firstly
$P = R^{-1} U^\dag$. Secondly, we can read off:
\begin{equation}
  \label{eq:37}
  F^\dag P^\dag P F = F^\dag U R^{-1} R^{-1} U^\dag F =  F^\dag \tilde{K} F
\end{equation}
So the \Kahler{} metric is equal to $P^\dag P$.

The important point to note is that in canoncally normalising, we have redefined our
superfields, so we must also redefine them in our superpotential. 
This is discussed in the next section.

\subsection{Yukawa sector after canonical normalisation}
\label{sec:quark-sector-after}

In this section we return to the important question of the form of the
Yukawa matrices in the correct canonically normalised basis.
In order to do this we 
would have to calculate the shifting in all of the fields in
the superpotential.  Unfortunately, algebraically diagonalising the
sub-block $\tilde{K}_\phi$ is intractable, even for such a simple
model. We therefore make a second assumption and neglect the
effects of canonical normalisation arising from this sector,
although we shall correctly consider the effects of canonical
normalisation arising from all the other sectors.

Even making this assumption, 
the expressions we get are not especially pleasant. We then substitute
in the form of the \vevs{} (eq.~(\ref{eq:vevs}) and
eq.~(\ref{eq:vevs_as_epsilons})). Having done this, we then expand the
cofactors of $F^i \Fbar^j H$ as a power series in $\epsilon$ around
the point $\epsilon = 0$.
The cofactors of $\epsilon^n$ are quite complicated, so we only
write out here the expression for the effective Yukawa for the
22 element. The full expressions for all four elements are
listed in Appendix \ref{sec:full_cn_yukawas}.
\begin{equation}
  \label{eq:12}
  Y_{23} = -a_1 \frac{b_3}{\sqrt{b_0 c_0} b_1}\epsilon^2 
  + a_2 \frac{1}{\sqrt{b_0 c_0 d_4}}\epsilon^3 + a_1 \frac{b_3(b_2 c_0 + b_0 (c_1 + c_2))}{2 b_0^{3/2} c_0^{3/2} b_1} \epsilon^3
  + \mathcal{O}(\epsilon^4)
\end{equation}
The important point to note is that, compared to
the 23 element of Eq.\ref{eq:1}, the leading order expansion in $\epsilon$
has changed. No longer is it at $\epsilon^3$, it is now
$\epsilon^2$. 

Note that we can write the expressions for
the canonically normalized off-diagonal Yukawa matrix elements $Y_{23}$ and
$Y_{32}$ in such a way that they would transform into each other
if we interchange $b_i \leftrightarrow c_i$, as would be expected.
We also note that the diagonal matrix elements
would transform into themselves under the same substitution, $b_i
\leftrightarrow c_i$.
This has been checked explicitly to the order in the Taylor
expansion shown in the Appendix.

Setting the $\mathcal{O}(1)$ parameters $b_i$,$c_i$ and $d_i$ to
unity, the Yukawa matrix then takes the canonical form:
\begin{equation}
  \label{eq:13}
  Y_c \sim 
  \left(
    \begin{array}{cc}
      (a_1+a_2)\epsilon^3 & -a_1 \epsilon^2 + (1.5a_1 + a_2) \epsilon^3 \\
      -a_1 \epsilon^2 + (1.5a_1 + a_2) \epsilon^3 & a_1 \epsilon -2 a_1 \epsilon^2 + a_2 \epsilon^3
    \end{array}
  \right)
  + \mathcal{O}(\epsilon^{4})
\end{equation}
We emphasise again that Eq.\ref{eq:13}
has a different power structure in $\epsilon$ to the
original, non-canonically normalised Yukawa in eq.~(\ref{eq:1}).

What has happened is that the unitary matrix which redefines our
fields has mixed them amongst themselves. This leads to a similar (but
different) high energy Yukawa texture. This certainly could be a
sufficiently different set-up to ruin any predictions that the
non-canonical model was designed to make. However we emphasise that
this result applies to the simplified $\SU{3}_F$ model
with a single expansion parameter, and not the 
realistic $\SU{3}_F$ model of
King and Ross \cite{King:2003rf} with two different expansion
parameters.

By comparing the non-canonical Yukawa matrix in eq.~(\ref{eq:1}) to
the canonical Yukawa matrix in eq.~(\ref{eq:13}), we can see that the
\Kahler{} mixing angles are large, of $\mathcal{O}(\epsilon)$.  
In the appendix, we have an expression
for the inverse P-matrix, $P^{-1}$. The large mixing effect can come
only from the mixing part of the transformation.
Schematically, the appearance of the $\epsilon^2$ leading order
terms in the off-diagonal elements can then be understood 
by neglecting all the
coefficients of $\mathcal{O}(1)$, as follows:
\begin{equation}
  \label{eq:39}
  Y_c \sim 
  \left(
    \begin{array}{cc}
      1 & \epsilon \\
      \epsilon & 1
    \end{array}
  \right)
  \left(
  \begin{array}{cc}
    \epsilon^3 & \epsilon^3 \\
    \epsilon^3 & \epsilon
  \end{array}
  \right)
  \left(
    \begin{array}{cc}
      1 & \epsilon \\
      \epsilon & 1
    \end{array}
  \right)
  \sim 
  \left(
    \begin{array}{cc}
      \epsilon^3 & \epsilon^2 \\
      \epsilon^2 & \epsilon
    \end{array}
  \right)
\end{equation}
which accounts for the appearance of the $\epsilon^2$ leading order
terms in the off-diagonal elements.

\subsection{Three generations of matter}
\label{sec:three-generations}

The procedure we have discussed for the second and third families
can straightforwardly be generalised to include also the first family
or indeed to any number of generations. 
The first thing to do is to write down all of the symmetries of the
model. Having done this, write down all of the non-renormalisable
operators up to the the chosen order in the UV cutoff, $M$.  In the
two generation case, this was to $\mathcal{O}(M^{-3})$. 
The next step is to write down the \Kahler{} potential
consistent with all the symmetries of the model, up to the same order
in the UV cutoff $M$ as the superpotential was expanded to.  For
tractability, some terms may have to be switched off to make the
\Kahler{} metric block diagonal as in eq.~(\ref{eq:8}). At this point,
the fields which break the family symmetry are replaced by their
\vevs.

Then one must find the matrices which canonically normalise each
sub-block of the \Kahler{} metric.  These will take the form of a
unitary matrix which diagonalises the sub-block, and then a rescaling
which takes it to the identity matrix of the appropriate size. Having
done this, the unnormalised fields can be written in terms of the
canonically normalised fields. If $\tilde{P}_S$ is the matrix which
diagonalises the sub-block $\tilde{K}_S$, and $\psi_S$ and
$\psi_S^\prime$ are respectively the unnormalised and canonically
normalised fields in the sub-block, then:
\begin{equation}
  \label{eq:14}
  \tilde{P}_S \tilde{K}_S \tilde{P}_S^\dag = \mathbf{1}
\end{equation}
\begin{equation}
  \label{eq:23}
  \psi_S = \tilde{P}_S \psi_S^\prime
\end{equation}

We then substitute eq.~(\ref{eq:23}) into the superpotential. Once we
have done this, the canonically normalised Yukawa matrix will be the
coefficient of $F^\prime \overline{F}^\prime H^\prime$.  At this
point, the Yukawa matrix elements may well be of the form of one
polynomial in expansion parameters, ( $\epsilon$ in the example model
) divided by another. In this case, to understand the power structure
in the expansion parameter, it is necessary to use a Taylor expansion
to get a power series in the expansion parameters ( we may do this
because the expansion parameters must be small in order for the whole
technique of non-renormalisable operators to work in the first place).

Having completed this, the end result is canonically normalised
three-generation Yukawa matrices, as required. Note that any step of
this calculation could in principle be intractable, and therefore some
simplifying assumptions may have to be made.

\section{Canonical origin of mixing angles}
\label{sec:interesting-possibility}

It is possible in principle that all fermion mixing angles
could originate from diagonal Yukawa couplings, via
canonical normalisation. To illustrate the idea, 
consider a two generation model, in which the non-canonical Yukawa
is diagonal, with the 33 element dominating over the 22 element:
\begin{equation}
  \label{eq:31}
  Y_{\mathrm{n.c.}} = 
  \left(
    \begin{array}{cc}
      \overline{\epsilon} & 0 \\
      0 & \lambda_t
    \end{array}
  \right)
\end{equation}
So $\overline{\epsilon} \ne 0$ and $\overline\epsilon \ll \lambda_t$
In general, the mixing part of the canonical normalisation can be
parameterised by a unitary rotation matrix, $U$, and the rescaling can
be parameterised by a diagonal matrix, $R$:
\begin{equation}
  \label{eq:32}
  Y_{c} = 
  \left(
    \begin{array}{cc}
      r_1 & 0 \\
      0 & r_1 
    \end{array}
  \right)
  \left(
    \begin{array}{cc}
      \cos\theta & -\sin\theta \\
      \sin\theta & \cos\theta
    \end{array}
  \right)
  \left(
    \begin{array}{cc}
      \overline{\epsilon} & 0 \\
      0 & \lambda_t
    \end{array}
  \right)
  \left(
    \begin{array}{cc}
      \cos\theta & \sin\theta \\
      -\sin\theta & \cos\theta
    \end{array}
  \right)
  \left(
    \begin{array}{cc}
      r_1 & 0 \\
      0 & r_2
    \end{array}
  \right)
\end{equation}
This leads to a canonical $Y$, $\phi = -\theta$
\begin{equation}
  \label{eq:33}
  Y_c = 
  \left(
    \begin{array}{cc}
      r_1^2(\overline\epsilon\cos^2_\phi + \lambda_t \sin^2\phi) &
      r_1r_2 \frac{\lambda_t - \overline{\epsilon}}{2} \sin 2\phi \\
      r_1r_2 \frac{\lambda_t - \overline{\epsilon}}{2} \sin 2\phi &
      r_2^2 (\overline\epsilon \sin^2\phi + \lambda_t \cos^2_\phi)
    \end{array}
  \right)
\end{equation}
Now consider the values for the parameters that $\sin\phi \approx \epsilon$,
$\overline{\epsilon} \approx \epsilon^n$ with $n > 3$,
$r_1 \approx r_2 \approx 1$ and $\lambda_t \approx \epsilon$:
\begin{equation}
  \label{eq:27}
  Y_c \approx \left(
    \begin{array}{cc}
      \epsilon^3 + \epsilon^n(1 - \epsilon^2) & \epsilon^2(1 - \epsilon^{n-1} ) \\
      \epsilon^2(1 - \epsilon^{n-1}) & \epsilon(1 - \epsilon^2) + \epsilon^{n+2} 
    \end{array}
    \right)
\end{equation}

By taking the leading order in $\epsilon$ and the leading two orders in $\epsilon$
in the 33 element, we can get a Yukawa matrix, post canonical-normalisation:

\begin{equation}
  \label{eq:36}
  Y_c \approx \left(
    \begin{array}{cc}
      \epsilon^3 & \epsilon^2 \\
      \epsilon^2 & \epsilon - \epsilon^3
    \end{array}
    \right)
\end{equation}
This look remarkably like the Yukawa matrix in the full case {\em before} canonical normalisation,
( eq. (\ref{eq:1})).

\section{Conclusions}
\label{sec:conclusions}

We have highlighted the important r\^{o}le that 
canonical normalisation of kinetic terms in flavour models 
based on family symmetries can play
in determining the Yukawa matrices. Even though the
kinetic terms may be correctly canonically normalised
to begin with, we have shown that they will inevitably be driven into 
a non-canonical form by a similar operator expansion
to that which determines the Yukawa operators.
Therefore in models based on family symmetry
canonical re-normalisation is mandatory before the
physical Yukawa matrices can be extracted.

In SUSY models with family symmetry, the \Kahler{} potential
should be considered to the same order in the UV cutoff as one takes
in the superpotential. Having done so, the \Kahler{} metric, which
follows from the \Kahler{} potential should be canonically normalised.
This will save the model from dangerous off-diagonal scalar mass
mixing terms in the super-CKM basis (and its leptonic analogue),
but the fields appearing in the superpotential must be redefined
leading to modified predictions for Yukawa matrices.

We have performed an explicit calculation of such mixing associated
with canonical normalisation of the \Kahler{} metric in a
supersymmetric model based on $SU(3)$ family symmetry,
and shown that such effects can significantly change the form of the
Yukawa matrix. In the simplified example considered,
one off-diagonal Yukawa element loses one power of an expansion parameter,
$\epsilon \approx \frac{1}{10}$, corresponding to that element growing
by an order of magnitude. We emphasise that this result does not
imply that the full realistic $\SU{3}_F$ model of
King and Ross \cite{King:2003rf} with two different expansion
parameters is incorrect. The analysis of the realistic $\SU{3}_F$ model
model with two different expansion parameters is more subtle, and 
such models may remain completely viable after canonical
normalisation \cite{GG}. 

We have also pointed out that the canonical form of the scalar mass
matrices takes a universal form as a direct result of the
simple supersymmetry breaking mechanism we have assumed.
The effects of canonical normalisation
on the scalar mass matrices in such realistic $\SU{3}_F$ models
recently considered in \cite{Ross:2004qn} must therefore also be
reconsidered \cite{GG}.

Finally we have pointed out that in principle quark mixing 
could originate entirely from canonical normalisation,
with only diagonal Yukawa couplings before canonical normalisation.
Although we have only considered a two family example explicitly,
we have indicated how the procedure generalises to the full three family 
case.

In conclusion, when looking at the flavour problem in effective field
theories based on family symmetries, it is not enough just to find
operators which gives a viable Yukawa structure. It is also necessary
to examine the structure of the kinetic terms, and ensure that the
Yukawa structure remains viable after canonically normalising the
kinetic terms, which redefines the fields.

\vskip 0.1in
\noindent
{\large {\bf Acknowledgements}}\\ I.P. thanks PPARC for a
studentship. We are also grateful to S.Antusch for useful
discussions. S.F.K. would like to thank L.Everett and G.Kane for
emphasising the importance of the \Kahler{} potential. We are
extremely grateful to Graham Ross for noticing some important errors
in an earlier version of this paper. We are
also grateful to T.Dent for a useful comment.

\appendix

\section{Expressions for the canonically normalised Yukawa elements,
and $P^{-1}_{LH}$}
\label{sec:full_cn_yukawas}

We write here the full expressions for the four Yukawa elements.

\begin{eqnarray}
\label{eq:15}
  Y_{22} &=& a_2 \frac{1}{\sqrt{b_0 c_0 d_4}} \epsilon^3 +
    a_1 \frac{b_3 c_3}{\sqrt{b_0 c_0 } b_1 c_1 }\epsilon^3 + \mathcal{O}(\epsilon^4)
    \\
  \label{eq:16}
  Y_{23} &=& -a_1 \frac{b_3}{\sqrt{b_0 c_0} b_1}\epsilon^2 
  + a_2 \frac{1}{\sqrt{b_0 c_0 d_4}}\epsilon^3 + a_1 \frac{b_3(b_2 c_0 + b_0 (c_1 + c_2))}{2 b_0^{3/2} c_0^{3/2} b_1} \epsilon^3
  + \mathcal{O}(\epsilon^4)
  \\
  Y_{32} &=& -a_1 \frac{c_3}{\sqrt{b_0 c_0} c_1}\epsilon^2 + a_2 \frac{1}{\sqrt{b_0 c_0 d_4}} \epsilon^3
  + a_1 \frac{c_3 ( c_2 b_0 + c_0 ( b_1 + b_2 ))}{2 b_0^{3/2} c_0^{3/2} c_1 \sqrt{d_4}} \epsilon^3
  + \mathcal{O}(\epsilon^4)
  \label{eq:17}
  \\
  \nonumber
  Y_{33} &=& a_1 \frac{1}{\sqrt{b_0 c_0}}\epsilon + - a_1 \frac{c_0 ( b_1 + b_2 ) + b_0 ( c_1 + c_2 ) }{2 b_0^{3/2} c_0^{3/2}} \epsilon^2
  + a_2 \frac{1}{\sqrt{b_0 c_0 d_4}}\epsilon^3 \\
  && + a_1 \frac{1}{8 b_0^{5/2} c_0^{5/2}} (
    \frac{c_0^2 ( 3 b_1^4 + 6b_1^3 b_2 -4 b_0^2 b_3^2 + b_1^2 ( 3b_2^2 - 4 b_0 (b_3 +b_4 +b_6 ) ) )}{b_1^2}
  \\
  \nonumber
  &&
  +
    \frac{b_0^2 ( 3 c_1^4 + 6 c_1^3 c_2 -4 c_0^2 c_3^2 + c_1^2 ( 3c_2^2 - 4c_0 (c_3 + c_4 + c_6)))}{c_1^2}\\
    &&
      + 2 b_0 c_0 ( b_1 + b_2)(c_1 + c_2))\epsilon^3 + \mathcal{O}(\epsilon^4)
  \label{eq:18}
\end{eqnarray}

These follow from the expressions for the inverse P-matrix after it has been Taylor expanded
in $\epsilon$ to order $\epsilon^3$ around the point $\epsilon = 0$. The full expression
for the left-handed P-matrix is then, to sub-leading order in $\epsilon$:
\begin{equation}
  \label{eq:40}
  P^{-1}_{LH} =
  \left(
    \begin{array}{cc}
      \frac{1}{\sqrt{b_0}} - \frac{b_2 \epsilon}{2 b_0^{3/2}} & \frac{b_3 \epsilon}{\sqrt{b_0} b_1 }
      - \frac{(b_1 + b_2) b_3 \epsilon^2}{2 b_0^{3/2} b_1 } \\
      - \frac{b_3 \epsilon}{\sqrt{b_0} b_1 } + \frac{b_2 b_3 \epsilon^2}{2 b_0^{3/2} b_1} &
      \frac{1}{\sqrt{b_0}} - \frac{(b_1 + b_2)\epsilon}{2 b_0^{3/2}}
    \end{array}
  \right)
\end{equation}

The structure of the right-handed equivalent is exactly the same, but with every $b_i$ replaced
with a $c_i$.

\end{document}